\documentclass[10pt,twocolumn]{IEEEtran}
\pdfoutput=1 
\usepackage[utf8]{luainputenc}
\usepackage{verbatim}
\usepackage{amsmath}
\usepackage{amssymb}
\usepackage{graphicx}
\usepackage{setspace}
\usepackage{wasysym}
\usepackage{esint}

\makeatletter
    
\usepackage{epsfig}\usepackage{amsfonts}\usepackage{latexsym}\usepackage{cite}\usepackage{array}\usepackage{amsmath}

\makeatother

\begin{document}

\title{Performance and Compensation of I/Q Imbalance in Differential STBC-OFDM }

\author{\IEEEauthorblockN{Lei Chen$^{\dagger}$, Ahmed G. Helmy$^{^{\star}}$,
Guangrong Yue$^{\dagger}$, Shaoqian Li$^{\dagger}$, and Naofal Al-Dhahir$^{^{\star}}$
{\small{}} }\\
{\small{}\IEEEauthorblockA{}$^{\dagger}${\small{} National Key
Laboratory of Communications, University of Electronic Science and
Technology of China, Chengdu 610054, China, lei.chen.2016@ieee.org,
\{yuegr, lsq\}@uestc.edu.cn}}\\
{\small{}\IEEEauthorblockA{$^{\star}$ The University of Texas at
Dallas, TX, USA, \{ahmed.g.helmy, aldhahir\}@utdallas.edu}}}
\maketitle
\begin{abstract}
Differential space time block coding (STBC) achieves full spatial
diversity and avoids channel estimation overhead. Over highly frequency-selective
channels, STBC is integrated with orthogonal frequency division multiplexing
(OFDM) to achieve high performance. However, low-cost implementation
of differential STBC-OFDM using direct-conversion transceivers is
sensitive to In-phase/Quadrature-phase imbalance (IQI). In this paper,
we quantify the performance impact of IQI at the receiver front-end
on differential STBC-OFDM systems and propose a compensation algorithm
to mitigate its effect. The proposed receiver IQI compensation works
in an adaptive decision-directed manner without using known pilots
or training sequences, which reduces the rate loss due to training
overhead. Our numerical results show that our proposed compensation
algorithm can effectively mitigate receive IQI in differential STBC-OFDM.
\vspace{-0.9cm}
\end{abstract}

\setlength{\textfloatsep}{0.3cm} \setlength{\floatsep}{0.3cm} \setlength{\abovecaptionskip}{0.1cm}

\begin{singlespace}

\section{Introduction }
\end{singlespace}

\begin{singlespace}
Space-time block-coded orthogonal frequency-division multiplexing
(STBC-OFDM) is an effective transceiver structure to mitigate the
wireless channel's frequency selectivity while realizing multipath
and spatial diversity gains \cite{ProceedingNaofal}. 
\end{singlespace}

To acquire channel knowledge for signal detection at the receiver,
STBC-OFDM schemes require the transmission of pilot symbols \cite{Tarokh2000}.
However, to avoid the rate loss due to pilot signal overhead, we may
want to forego channel estimation in order to reduce the increased
cost of channel estimation and the degradation of tracking quality
in a fast time-varying environment \cite{Diggavi2002,S_luDiff}. Differential
STBC transmission/detection achieves this goal and has been successfully
integrated with STBC-OFDM \cite{Diggavi2002,Tarokh2000,Hughes2000}.

\begin{singlespace}
Although a differential STBC-OFDM system avoids the overhead of channel
estimation, a low-cost transceiver implementation based on the direct-conversion
architecture suffers from analog/RF impairments. The impairments in
the analog components are mainly due to the uncontrollable fabrication
process variations. Since most of these impairments cannot be effectively
eliminated in the analog domain, an efficient compensation algorithm
in the digital baseband domain would be highly desirable for designing
such low-cost wireless transceivers. One of the main sources of the
analog components impairments is the imbalance between the In-phase
(I) and Quadrature-phase (Q) branches when the received radio-frequency
(RF) signal is down-converted to baseband. The I/Q imbalance (IQI)
arises due to mismatches between the I and Q branches from the ideal
case, i.e., from the exact $90^{o}$ phase difference and equal amplitudes
between the sine and cosine branches. In OFDM systems, IQI destroys
the subcarriers orthogonality by introducing inter-carrier interference
(ICI) between mirror subcarriers which can lead to serious performance
degradation \cite{Tarighat2005}. 

Several papers investigated IQI in single-input single-output OFDM
system (see \cite{Tarighat2005}\cite{Tarighat2007} and the references
therein). There are also several works dealing with IQI in coherent
multiple-antenna systems. In \cite{narasimhan2010reduced}, a super-block
structure for the Alamouti STBC scheme is designed to ensure orthogonality
in the presence of IQI. In \cite{Jointestimationtransmitter_Marey_2013},
an Expectation-Maximization-based algorithm is proposed to deal with
IQI in Alamouti-based STBC-OFDM systems. An equalization algorithm
is proposed to overcome IQI in STBC-OFDM systems in \cite{STBCMIMOOFDM_Tandur_2008}.
The authors in \cite{PerformanceAnalysisSpace_Zou_2007} analyze and
compensate IQI in single-carrier STBC systems. 

Although the compensation of IQI in STBC-OFDM systems is well studied,
all existing works deal with IQI either a) in coherent system, where
the channel state information (CSI) is known or estimated at the receiver;
or b) in a blind sense, where the signal is recovered by some statistical
method based on signal properties (such as constant-modulus or circularity)
\cite{Anttila2008}. \textbf{To the best of our knowledge, there is
no previous work dealing with IQI in differential transmission systems}.
Although blind estimation also does not require channel state information
for detection, the input symbols are not differentially encoded and
decoded and blind compensation algorithms suffer from local optima
and very slow convergence. In this paper, we analyze the impact of
the receiver IQI (RX-IQI) in differential STBC-OFDM (DSTBC-OFDM) systems
and propose an adaptive decision-directed scheme that compensates
for RX-IQI without knowing or estimating the channel information.
The rest of this paper is organized as follows

The system model of DSTBC-OFDM is developed in Section \ref{sec:We-consider-an}.
In Section \ref{sec:formulate}, we formulate the problem of RX-IQI
in DSTBC-OFDM and discuss the impact of RX-IQI on the bit error rate
(BER) performance of DSTBC-OFDM. We propose a decision-directed IQI
compensation algorithm in Section \ref{sec:compensateion} and the
numerical results are presented in Section \ref{sec:numerical}. Finally,
we conclude our paper in Section \ref{sec:conc}. 

\textit{Notations}: Unless further noted, matrices and vectors are
denoted by upper-case and lower-case boldface, respectively. We denote
the Hermitian, i.e. complex-conjugate transpose of a matrix or a vector
by ${{(\cdot)}^{H}}$. The conjugate and transpose of a matrix, a
vector, or a scalar is denoted by ${{(\cdot)}^{*}}$ and ${{(\cdot)}^{T}}$,
respectively. The symbol ${{[\mathbf{A}]}_{m,n}}$ denotes the entry
at the $m$-th row and the $n$-th column of matrix $\mathbf{A}$.
Matrix $\mathbf{F}$ is the $N$-point Discrete Fourier Transform
(DFT) matrix whose entries are given by: ${{[\mathbf{F}]}_{m,n}}=\frac{1}{\sqrt{N}}\exp(-j\frac{2\pi}{N}mn)$,
with $0\le m,n\le N-1$. $\textrm{Re}\{\cdot\}$ and $\textrm{Im}\{\cdot\}$
denote the real and image parts of a complex number, respectively.
\vspace{0.3cm}
\end{singlespace}

\begin{singlespace}

\section{\label{sec:We-consider-an}System Model }
\end{singlespace}

\begin{spacing}{1.1}
We consider an Alamouti-based STBC-OFDM wireless communication system
equipped with two transmit antennas and a single receive antenna.
In STBC-OFDM systems, the modulated sequence is divided into $K$
blocks of a length of $2N$ symbols (where $N$ is a power of $2$)
encoded by a space-time block encoder to be transmitted over the $\left(2k+1\right)$-th
and $\left(2k+2\right)$-th OFDM symbols ($k=0,\cdots,K-1$). We choose
the $2\times2$ Alamouti STBC structure \cite{Alamouti1998} as the
core STBC due to its orthogonality advantages. In Alamouti STBC-OFDM,
two complex information symbols ($x_{1}$ and $x_{2}$) drawn from
a constant modulus signal constellation $\varPsi$ are simultaneously
transmitted from two transmit antennas for the data block index $k$
over the $\left(2k+1\right)$-th and $\left(2k+2\right)$-th OFDM
symbols.
\end{spacing}

For each two consecutive OFDM symbols, we define the $2\times1$ vectors
$\mathbf{u}_{k}^{\left(1\right)}\left(n\right)$ and $\mathbf{u}_{k}^{\left(2\right)}\left(n\right)$
to be the information vectors at the $n$-th subcarrier ($n=1,\cdots,N$)
over the $\left(2k+1\right)$-th and $\left(2k+2\right)$-th OFDM
symbols, respectively. Hence, the $2\times2$ Alamouti STBC information
matrix $\mathbf{U}_{k}(n)$ at the $n$-th subcarrier over the $\left(2k+1\right)$-th
and $\left(2k+2\right)$-th OFDM symbols is given by $\mathbf{U}_{k}(n)\triangleq\left[\mathbf{u}_{k}^{\left(1\right)}\left(n\right)\:\brokenvert\:\mathbf{u}_{k}^{\left(2\right)}\left(n\right)\right]$
whose columns, $\mathbf{u}_{k}^{\left(1\right)}\left(n\right)$ and
$\mathbf{u}_{k}^{\left(2\right)}\left(n\right)$, are defined as follows{\small{}
\begin{align}
\mathbf{u}_{k}^{\left(1\right)}\left(n\right) & =\left[\begin{array}{cc}
x_{1} & -x_{2}^{*}\end{array}\right]^{T}\mathbf{u}_{k}^{\left(2\right)}\left(n\right)=\left[\begin{array}{cc}
x_{2} & x_{1}^{*}\end{array}\right]^{T}
\end{align}
}where the columns of $\mathbf{U}_{k}(n)$ correspond to the information
data to be transmitted from the two transmit antennas over the $\left(2k+1\right)$-th
and $\left(2k+2\right)$-th OFDM symbols, respectively. Then, the
output of the STBC encoder block is further passed through a serial-to-parallel
converter producing two streams of data blocks of length $N$, with
$N$ being the number of subcarriers. Each block is applied to a per-stream
$N$-point Inverse Discrete Fourier Transform (IDFT). To avoid intersymbol
interference (ISI), a cyclic prefix (CP) of length $\nu$ is added
to the $N$ time-domain IDFT-output samples, resulting in an OFDM
symbol of length $N+\nu$. We model the frequency-selective fading
channel between the $i$-th transmit antenna and the single receive
antenna as a finite impulse response (FIR) filter with $L+1$ independent
taps $i.e.$ {\small{}$\mbox{ }\mathbf{h}_{i}=[h_{0}^{\left(i\right)}\,h_{1}^{\left(i\right)}\,\cdots\,h_{L}^{\left(i\right)}]{}^{T}$}
where $L\leq\nu$ to preserve the orthogonality between the subcarriers. 

Without loss of generality, we consider the block index $k=0$ transmitted
over the $1^{st}$ and $2^{nd}$ OFDM symbols, respectively, hence,
we omit the index $k$ for simplicity. However, the following system
model can be applied to any data block index $k$ corresponding to
the $\left(2k+1\right)$-th and $\left(2k+2\right)$-th OFDM symbols.
After removing the CP at the receiver, the $N\times1$ time-domain
received vectors corresponding to the $1^{st}$ and $2^{nd}$ OFDM
symbols from all $N$ subcarriers at the receive antenna, denoted
by $\mathbf{y}_{1}$ and $\mathbf{y}_{2}$, are given by {\small{}
\begin{align}
\mathbf{y}_{1} & =\mathbf{H}_{1}^{c}\mathbf{s}_{1}+\mathbf{v}_{1}\qquad,\qquad\mathbf{y}_{2}=\mathbf{H}_{2}^{c}\mathbf{s}_{2}+\mathbf{v}_{2}\label{eq:model-1}
\end{align}
}where $\mathbf{H}_{1}^{c}$, $\mathbf{H}_{2}^{c}$, $\mathbf{s}_{1}$,
and $\mathbf{s}_{2}$ are defined as follows{\small{}
\begin{equation}
\mathbf{H}_{1}^{c}=\left[\mathbf{H}_{1,1}^{c}\,,\,\mathbf{H}_{1,2}^{c}\right]\thinspace,\thinspace\mathbf{H}_{2}^{c}=\left[\mathbf{H}_{2,1}^{c}\,,\,\mathbf{H}_{2,2}^{c}\right]
\end{equation}
} {\small{}\vspace{-0.5cm}
\begin{align}
\mathbf{s}_{1} & =\left[\begin{array}{ccccc}
s_{1,1}^{\left(1\right)} & \cdots & s_{n,1}^{\left(1\right)} & \cdots & s_{N,1}^{\left(1\right)}\end{array}\right.\nonumber \\
 & \left.\begin{array}{ccccc}
\quad s_{1,1}^{\left(2\right)} & \cdots & s_{n,1}^{\left(2\right)} & \cdots & s_{N,1}^{\left(2\right)}\end{array}\right]^{T}\\
\mathbf{s}_{2} & =\left[\begin{array}{ccccc}
s_{1,2}^{\left(1\right)} & \cdots & s_{n,2}^{\left(1\right)} & \cdots & s_{N,2}^{\left(1\right)}\end{array}\right.\nonumber \\
 & \left.\begin{array}{ccccc}
\mathbf{\quad}s_{1,2}^{\left(2\right)} & \cdots & s_{n,2}^{\left(2\right)} & \cdots & s_{N,2}^{\left(2\right)}\end{array}\right]^{T}
\end{align}
}where $\mathbf{H}_{1,i}^{c}$ and $\mathbf{H}_{2,i}^{c}$ ($i=1,2$)
are the $N\times N$ circulant time-domain channel impulse response
matrix corresponding to the channel from the $i$-th transmit antenna
over the $1^{st}$ and $2^{nd}$ OFDM symbols, respectively. Assuming
a quasi-static channel model over the $1^{st}$ and $2^{nd}$ OFDM
symbols, the first row of $\mathbf{H}_{1,i}^{c}$ and $\mathbf{H}_{2,i}^{c}$
is the same and given by {\small{}$\left[h_{0}^{\left(i\right)}\,h_{1}^{\left(i\right)}\,\cdots\,h_{L}^{\left(i\right)}\,0\,\cdots\,0\right]$},
hence, {\small{}$\mathbf{H}_{1,i}^{c}=\mathbf{H}_{2,i}^{c}=\mathbf{H}_{i}^{c}$}.
Moreover, $\mathbf{s}_{1}$ and $\mathbf{s}_{2}$ are the $2N\times1$
time-domain transmitted vectors from the two transmit antennas corresponding
to the $1^{st}$ and $2^{nd}$ OFDM symbols, respectively. In addition,
$s{}_{n,1}^{\left(i\right)}$ and $s_{n,2}^{\left(i\right)}$ are
the transmitted signals from the $i$-th transmit antenna forming
the $2\times1$ transmitted signal vectors $\mathbf{s}_{n}^{\left(1\right)}$
and $\mathbf{s}_{n}^{\left(2\right)}$ at the $n$-th subcarrier over
the $1^{st}$ and $2^{nd}$ OFDM symbols, respectively.

\begin{spacing}{1.2}
Moreover, $\mathbf{v}_{1}$ and $\mathbf{v}_{2}$ are the zero-mean
time-domain additive white Gaussian noise (AWGN) vectors, whose elements
are mutually independent with covariance matrix ${{\sigma}^{2}}\mathbf{I}_{N\times1}$.
Since, $\mathbf{H}_{i}^{c}$ is a quasi-static channel that can be
diagonalized by the $N$-point Discrete Fourier Transform (DFT) matrix
as $\mathbf{H}_{i}^{c}=\mathbf{F}\mathbf{\Lambda}_{i}\mathbf{F}^{H}$,
where $\mathbf{\Lambda}_{i}=diag\left\{ \lambda_{i}\right\} $ and
$\lambda_{i}$ is the $N\times1$ vector corresponding to the channel
coefficients from the $i$-th transmit antenna and is given by {\small{}
\begin{equation}
\lambda_{i}=\sqrt{N}\mathbf{F}^{H}\left[\begin{matrix}\mbox{ }\mathbf{h}_{i}\\
\mathbf{0}_{\left(N-\left(L+1\right)\right)\times1}
\end{matrix}\right]\label{eq:lmda}
\end{equation}
}where $\mathbf{F}$ is the $N$-point DFT matrix and ${{\mathbf{0}}_{(N-\left(L+1\right))\times1}}$
is a zero vector of length $N-\left(L+1\right)$. After applying the
$N$-point DFT at the receiver to the time-domain received vectors
$\mathbf{y}_{1}$ and $\mathbf{y}_{2}$ in Eq. (\ref{eq:model-1}),
the $2\times1$ frequency-domain received vectors $\mathbf{z}_{1}=\mathbf{F}\mathbf{y}_{1}$
and $\mathbf{z}_{2}=\mathbf{F}\mathbf{y}_{2}$ are given by\vspace{-0.2cm}{\small{}
\begin{align}
\mathbf{z}_{1} & =\left[\begin{array}{cc}
\mathbf{\Lambda}_{1} & 0\\
0 & \mathbf{\Lambda}_{2}
\end{array}\right]\mathbf{\underline{s}}_{1}+\mathbf{\underline{v}}_{1}\;,\;\mathbf{z}_{2}=\left[\begin{array}{cc}
\mathbf{\Lambda}_{1} & 0\\
0 & \mathbf{\Lambda}_{2}
\end{array}\right]\mathbf{\mathbf{\underline{s}}}_{2}+\mathbf{\underline{v}}_{2}
\end{align}
}where $\mathbf{\underline{v}}_{1}$ and $\mathbf{\underline{v}}_{2}$
are the $2N\times1$ frequency-domain additive noise vectors. In addition,
$\underline{\mathbf{s}}_{1}$ and $\mathbf{\underline{s}}_{2}$ are
the $2N\times1$ frequency-domain transmitted vectors from the two
transmit antennas corresponding to the $1^{st}$ and $2^{nd}$ OFDM
symbols, respectively. Hence, the $2\times2$ combined frequency-domain
received data matrix at the $n$-th subcarrier constructed in the
Alamouti STBC matrix form, $\mathbf{Z}\left(n\right)$, is given by
\cite{Diggavi2002}{\small{}
\begin{align}
\underset{\mathbf{Z}\left(n\right)}{\underbrace{\left[\begin{array}{cc}
z_{1}(n) & z_{2}(n)\\
-z_{2}^{*}(n) & z_{1}^{*}(n)
\end{array}\right]}} & =\underset{\mathbf{\Lambda}\left(n\right)}{\underbrace{\left[\begin{matrix}{{\lambda}_{1}}(n) & {{\lambda}_{2}}(n)\\
-{\lambda_{2}^{*}}{(n)} & {\lambda_{1}^{*}}{(n)}
\end{matrix}\right]}}\underset{\mathbf{S}\left(n\right)}{\underbrace{\left[\begin{array}{cc}
\underline{s}_{1}(n) & \underline{s}_{2}(n)\\
-\underline{s}_{2}^{*}(n) & \underline{s}_{1}^{*}(n)
\end{array}\right]}}\nonumber \\
+ & \underset{\mathbf{V}\left(n\right)}{\underbrace{\left[\begin{array}{cc}
\underline{v}_{1}(n) & \underline{v}_{2}(n)\\
-\underline{v}_{2}^{*}(n) & \underline{v}_{1}^{*}(n)
\end{array}\right]}}\label{eq:model2}\\
\Rightarrow\mathbf{Z}\left(n\right) & =\mathbf{\Lambda}\left(n\right)\mathbf{S}\left(n\right)+\mathbf{V}\left(n\right)\label{eq:model2-1}
\end{align}
}where $z_{1}(n)$, $z_{2}(n)$, ${{\lambda}_{1}}(n)$, ${{\lambda}_{2}}(n)$,
$\underline{s}_{1}(n)$, $\underline{s}_{2}(n)$, $\underline{v}_{1}(n)$,
and $\underline{v}_{2}(n)$ are the $n$-th subcarrier component of
the vectors $\mathbf{z}_{1}$, $\mathbf{z}_{2}$, ${{\lambda}_{1}}$,
${{\lambda}_{2}}$, $\mathbf{\underline{s}}_{1}$, $\mathbf{\underline{s}}_{2}$,
$\mathbf{\underline{v}}_{1}$, and $\mathbf{\underline{v}}_{2}$,
respectively. 
\end{spacing}

In general, for data blocks indices $k$ and $k+1$, we assume that
the STBC-modulated information for the $n$-th subcarrier is differentially-encoded
over the time domain (DSTBC-OFDM). These two consecutive data blocks
correspond to the $\left(2k+1\right)$, $\left(2k+2\right)$, $\left(2\left(k+1\right)+1\right)$,
and $\left(2\left(k+1\right)+2\right)$ four consecutive OFDM symbols.
Assuming that the channel is quasi-static over four consecutive OFDM
symbols, the modified system model in Eq. (\ref{eq:model2-1}) adopting
the differential encoding corresponding to these four consecutive
OFDM symbols can be formulated as follows \cite{Diggavi2002}
\begin{equation}
\mathbf{S}_{k+1}(n)=\mathbf{S}_{k}(n)\mathbf{U}_{k+1}(n)\label{difsu}
\end{equation}
where $\mathbf{S}_{k}(n)$ is the $2\times2$ STBC transmitted data
matrix corresponding to the $\left(2k+1\right)$-th, $\left(2k+2\right)$-th
OFDM symbols of the $k$-th data block at the $n$-th subcarrier defined
in Eq. (\ref{eq:model2}). Similarly, $\mathbf{S}_{k+1}(n)$ and $\mathbf{U}_{k+1}(n)$
are the $2\times2$ STBC transmitted data matrix and information matrix,
respectively, corresponding to the $\left(2\left(k+1\right)+1\right)$-th
and $\left(2\left(k+1\right)+2\right)$-th OFDM symbols of the $\left(k+1\right)$-th
data block at the $n$-th subcarrier. Based on the differential encoding
in Eq. (\ref{difsu}), $\mathbf{Z}_{k}\left(n\right)$ and $\mathbf{Z}_{k+1}\left(n\right)$
are defined by
\begin{align}
\mathbf{Z}_{k}\left(n\right) & =\mathbf{\Lambda}(n)\mathbf{S}_{k}\left(n\right)+\mathbf{V}_{k}\left(n\right)\\
\mathbf{Z}_{k+1}\left(n\right) & =\mathbf{\Lambda}(n)\mathbf{S}_{k+1}\left(n\right)+\mathbf{V}_{k+1}\left(n\right)\nonumber \\
 & =\mathbf{\Lambda}(n)\mathbf{S}_{k}(n)\mathbf{U}_{k+1}(n)+\mathbf{V}_{k+1}\left(n\right)\label{eq:freqDiff}
\end{align}

The maximum likelihood (ML) decoder for the information matrix $\mathbf{U}_{k+1}(n)$
is given by\cite{Hughes2000} 
\begin{equation}
\mathbf{\hat{U}}_{k+1}(n)=\underset{\mathbf{\quad\quad\quad U}_{k+1}(n)}{\mathop{\arg}\quad\mathop{\max}}\,\left\{ \mathbf{U}_{k+1}^{H}(n)\mathbf{Z}_{k}^{H}\left(n\right)\mathbf{Z}_{k+1}\left(n\right)\right\} \label{ML}
\end{equation}
where $\mathbf{U}_{k+1}(n)$ is chosen from the Alamouti matrix sets
formed by all possible information matrices. \vspace{0.14cm}

\begin{spacing}{1.3}

\section{\label{sec:formulate}DSTBC-OFDM under Receiver I/Q Imbalance (RX-IQI) }
\end{spacing}

\vspace{-0.2cm}We adopt the time-domain receiver RX-IQI model defined
in \cite{Anttila2008} where the time-domain signal $b'\left(t\right)$
distorted by the RX-IQI is modeled as follows
\begin{equation}
b'\left(t\right)=\alpha_{r}b\left(t\right)+\beta_{r}b^{*}\left(t\right)\label{eq:iqi}
\end{equation}
 where $b\left(t\right)$ is the IQI-free received signal and the
parameters $({{\alpha}_{r}},{{\beta}_{r}})$ are RX-IQI parameters,
and they are defined by {\small{}
\begin{align}
\alpha_{r} & =\frac{1}{2}(1+g_{r}e^{-j\phi_{r}})\ ,\ \beta_{r}=\frac{1}{2}(1-g_{r}e^{j\phi_{r}})
\end{align}
}where $\phi_{r}$ and $g_{r}$ are the phase and the amplitude imbalance
between the I and Q branches. The amplitude imbalance is often denoted
in dB as ${{\kappa}_{r}}(dB)=20\log(g_{r})$. The overall imbalance
of a receiver is measured by the Image Rejection Ratio (IRR), which
is defined by $\mbox{IRR}(dB)\triangleq-10\log_{10}(\rho)=-10\log_{10}(|{{\beta}_{r}}|^{2}/|{{\alpha}_{r}}|^{2})=-20\log_{10}(|{{\beta}_{r}}|/|{{\alpha}_{r}}|)$.

Based on the RX-IQI model in Eq. (\ref{eq:iqi}), the time-domain
received signal vector $\mathbf{y}$ after RX-IQI will be transformed
into the distorted signal $\mathbf{y}'$ given by \vspace{-0.2cm}

{\small{}
\begin{equation}
\mathbf{y}'=\alpha_{r}\mathbf{y}+\beta_{r}\mathbf{y}^{*}\label{eq:iqimodel}
\end{equation}
}{\small \par}

Discarding the samples corresponding to the first and $\left(\frac{N}{2}+1\right)$
subcarriers, the effect of the RX-IQI on the $n$-th subcarrier of
the DSTBC-OFDM received signal is basically introducing ICI from its
$\left(N-n+2\right)$ image subcarrier \cite{Tarighat2007}. Based
on the DSTBC-OFDM model in Eq. (\ref{eq:freqDiff}), the frequency-domain
RX-IQI-distorted received signals $\mathbf{Z'}_{k}\left(n\right)$
and $\mathbf{Z'}_{k+1}\left(n\right)$ are given by\vspace{-0.5cm}

\begin{align}
\mathbf{\mathbf{Z'}}_{k}\left(n\right) & =\mathbf{A}_{r}\left(\mathbf{\Lambda}(n)\mathbf{S}_{k}\left(n\right)+\mathbf{V}_{k}\left(n\right)\right)\nonumber \\
 & +\mathbf{B}_{r}\left(\mathbf{\bar{\Lambda}}(n)\mathbf{\bar{S}}_{k}\left(n\right)+\mathbf{\bar{V}}_{k}\left(n\right)\right)\label{eq:freDiff2}\\
\mathbf{\mathbf{Z'}}_{k+1}\left(n\right) & =\mathbf{A}_{r}\left(\mathbf{\Lambda}(n)\mathbf{S}_{k+1}(n)+\mathbf{V}_{k+1}\left(n\right)\right)\nonumber \\
 & +\mathbf{B}_{r}\left(\mathbf{\bar{\Lambda}}(n)\mathbf{\bar{S}}_{k+1}(n)+\mathbf{\bar{V}}_{k+1}\left(n\right)\right)\nonumber \\
 & =\mathbf{A}_{r}\left(\mathbf{\Lambda}(n)\mathbf{S}_{k}(n)\mathbf{U}_{k+1}(n)+\mathbf{V}_{k+1}\left(n\right)\right)\nonumber \\
 & +\mathbf{B}_{r}\left(\mathbf{\bar{\Lambda}}(n)\mathbf{\bar{S}}_{k}(n)\mathbf{\bar{U}}_{k+1}(n)+\mathbf{\bar{V}}_{k+1}\left(n\right)\right)\label{eq:freqDiff-1}
\end{align}
where the matrices $\mathbf{A}_{r}$, $\mathbf{B}_{r}$ , $\mathbf{\bar{\Lambda}}(n)$,
$\mathbf{\bar{S}}_{k}\left(n\right)$, $\mathbf{\bar{S}}_{k+1}\left(n\right)$,
$\mathbf{\bar{V}}_{k}\left(n\right)$, $\mathbf{\bar{V}}_{k+1}\left(n\right)$,
and $\mathbf{\bar{U}}_{k+1}(n)$ are given by

\begin{singlespace}
{\small{}
\begin{align}
{{\mathbf{A}}_{r}} & =\left[\begin{matrix}{{\alpha}_{r}} & 0\\
0 & {{\alpha}_{r}}^{*}
\end{matrix}\right]\;,\;{{\mathbf{B}}_{r}}=\left[\begin{matrix}{{\beta}_{r}} & 0\\
0 & {{\beta}_{r}}^{*}
\end{matrix}\right]\label{eq:iqiparm}\\
\mathbf{\bar{\Lambda}}(n) & =\left[\begin{array}{cc}
{{\lambda}_{1}}(N-n+2) & {{\lambda}_{2}}(N-n+2)\\
-{\lambda_{2}^{*}}{(N-n+2)} & {\lambda_{1}^{*}}{(N-n+2)}
\end{array}\right]\\
\mathbf{\bar{S}}_{k}\left(n\right) & =\left[\begin{array}{cc}
\underline{s}_{2k+1}(N-n+2) & \underline{s}_{2k+2}(N-n+2)\\
-\underline{s}_{2k+2}^{*}(N-n+2) & \underline{s}_{2k+1}^{*}(N-n+2)
\end{array}\right]\\
\mathbf{\bar{V}}_{k}\left(n\right) & =\left[\begin{array}{cc}
\underline{v}_{2k+1}(N-n+2) & \underline{v}_{2k+2}(N-n+2)\\
-\underline{v}_{2k+2}^{*}(N-n+2) & \underline{v}_{2k+1}^{*}(N-n+2)
\end{array}\right]\\
\mathbf{\bar{V}}_{k+1}\left(n\right) & =\left[\begin{array}{cc}
\mathbf{\underline{v}}_{2k+3}(N-n+2) & \underline{v}_{2k+4}(N-n+2)\\
-\underline{v}_{2k+4}^{*}(N-n+2) & \underline{v}_{2k+3}^{*}(N-n+2)
\end{array}\right]\\
\mathbf{\bar{U}}_{k+1}(n) & =\left[\mathbf{u}_{k+1}^{\left(1\right)}\left(N-n+2\right)\:\brokenvert\:\mathbf{u}_{k+1}^{\left(2\right)}\left(N-n+2\right)\right]\\
\mathbf{S}_{k+1}(n) & =\mathbf{S}_{k}(n)\mathbf{U}_{k+1}(n)\\
\mathbf{\bar{S}}_{k+1}(n) & =\mathbf{\bar{S}}_{k}(n)\mathbf{\bar{U}}_{k+1}(n)
\end{align}
}where $\underline{s}_{2k+1}(N-n+2)$, $\underline{s}_{2k+2}(N-n+2)$,
$\underline{v}_{2k+1}(N-n+2)$, and $\underline{v}_{2k+2}(N-n+2)$
are the $\left(N-n+2\right)$ subcarrier component of the vectors
$\mathbf{\underline{s}}_{2k+1}$, $\mathbf{\underline{s}}_{2k+2}$,
$\mathbf{\underline{v}}_{2k+1}$, and $\mathbf{\underline{v}}_{2k+2}$,
respectively. \vspace{0.15cm}
\end{singlespace}

\begin{singlespace}

\subsection{\label{sub:ber}Performance Analysis of DSTBC-OFDM under RX-IQI }
\end{singlespace}

\begin{singlespace}
In this subsection, we analyze the impact of RX-IQI on an individual
subcarrier in DSTBC-OFDM with M-PSK signaling. We asymptotically quantify
the bit-error rate (BER) floor caused by RX-IQI and its corresponding
equivalent signal-to-noise ratio (SNR) compared to that of the IQI-free
system. 

For simplicity, we modify the definitions of the RX-IQI diagonal parameters'
matrices $\mathbf{A}_{r}$ and $\mathbf{B}_{r}$ to be $\mathbf{A}_{r}=\left|\alpha_{r}\right|\mathbf{I}$
and $\mathbf{B}_{r}=\left|\beta_{r}\right|\mathbf{I}$, respectively.
This modification is a valid assumption since the impact of the RX-IQI
is usually measured by IRR which basically depends on $\left|\alpha_{r}\right|$
and $\left|\beta_{r}\right|$. Moreover, the received signal usually
has a uniform phase distribution which makes the effect of the phase
of RX-IQI parameters irrelevant. For simplicity, we omit the subcarrier
index $n$. 

Based on these assumed modifications, the frequency-domain RX-IQI-distorted
received signals $\mathbf{Z'}_{k}$ and $\mathbf{Z'}_{k+1}$ can be
re-written as follows\vspace{-0.2cm}

\begin{align}
\mathbf{\mathbf{Z'}}_{k} & =|{{\alpha}_{r}}|\mathbf{\Lambda}{{\mathbf{S}}_{k}}+|{{\beta}_{r}}|\mathbf{\mathbf{\bar{\Lambda}}}{{\mathbf{\bar{S}}}_{k}}+|{{a}_{r}}|{{\mathbf{V}}_{k}}+|{{\beta}_{r}}|{{\mathbf{\bar{V}}}_{k}}\\
\mathbf{\mathbf{Z'}}_{k+1} & =|{{\alpha}_{r}}|\mathbf{\Lambda}{{\mathbf{S}}_{k}}\mathbf{U}_{k+1}+|{{\beta}_{r}}|\mathbf{\mathbf{\bar{\Lambda}}}{{\mathbf{\bar{S}}}_{k}}\mathbf{\mathbf{\bar{U}}}_{k+1}\nonumber \\
 & +|{{\alpha}_{r}}|\mathbf{V}_{k+1}+|{{\beta}_{r}}|\mathbf{\bar{V}}_{k+1}
\end{align}

From Eq. (\ref{ML}), the decoding metric for the ML decoder becomes
\begin{equation}
\begin{aligned}\mathbf{\mathbf{Z'}}_{k}^{H}\mathbf{\mathbf{Z'}}_{k+1} & =|\lambda|^{2}|{{\alpha}_{r}}{{|}^{2}}\mathbf{U}_{k+1}\\
 & +\underbrace{|{{\alpha}_{r}}{{\beta}_{r}}|\left({{\mathbf{S}}_{k}}^{H}{{\mathbf{\Lambda}}^{H}}\mathbf{\bar{\Lambda}{\mathbf{\bar{S}}}}_{k+1}+{{\mathbf{\bar{S}}}_{k}}^{H}{{\mathbf{\bar{\Lambda}}}^{H}}\mathbf{\Lambda{\mathbf{S}}}_{k+1}\right)}_{\mathbf{\Theta}}+{{\mathbf{V}}_{r}}
\end{aligned}
\label{eq:detectionmetric}
\end{equation}
where ${{\mathbf{V}}_{r}}=|{{\alpha}_{r}}{{|}^{2}}{{\mathbf{V}}_{k}}^{H}\mathbf{\Lambda{\mathbf{S}}}_{k}\mathbf{U}_{k+1}+|{{\alpha}_{r}}{{|}^{2}}{{\mathbf{S}}_{k}}^{H}{{\mathbf{\Lambda}}^{H}}\mathbf{V}_{k+1}$,
$|\lambda|^{2}\mathbf{I}=\mathbf{\Lambda}^{H}\mathbf{\Lambda}$, and
$|\bar{\lambda}|^{2}\mathbf{I}=\bar{\mathbf{\Lambda}}^{H}\bar{\mathbf{\Lambda}}$.
Recall that the differentially encoded matrices, ${{\mathbf{S}}_{k}}$
and $\mathbf{{\mathbf{\bar{S}}}}_{k}$, whose entries are sums of
numerous products of PSK symbols. To simplify the analysis and gain
more insights, we ignore the dependence between these PSK symbols
products. For a long input data sequence, we apply the central limit
theorem (CLT) to approximate the distributions of the entries of ${{\mathbf{S}}_{k}}$
and $\mathbf{{\mathbf{\bar{S}}}}_{k}$ by the two uncorrelated zero-mean
Gaussian distributions ${{\mathbf{S}}_{k}}\sim\mathcal{{N}}\left(0,\,\frac{1}{2}\right)$
and $\mathbf{{\mathbf{\bar{S}}}}_{k}\sim{N}\left(0,\,\frac{1}{2}\right)$
with a variance of $\frac{1}{2}$ to satisfy the power constraint
${{\mathbf{S}}_{k}}{{\mathbf{S}}_{k}}^{H}={{\mathbf{\bar{S}}}_{k}}{{\bar{\mathbf{S}}}_{k}}^{H}=\mathbf{I}$.

The detection of symbols in $\mathbf{U}_{k+1}$ is totally decided
by the detection metric in Eq.(\ref{eq:detectionmetric}). For a given
channel realization of the desired subcarrier $\mathbf{\Lambda}$
and image subcarrier $\mathbf{\bar{\Lambda}}$, the instantaneous
probability of error of the MPSK symbols in $\mathbf{U}_{k+1}$ is
decided by the instantaneous equivalent signal-to-interference-plus-noise
ratio (SINR) of $\mathbf{U}_{k+1}$ in the decoding metric for a given
channel realization\cite{Torabi2007}. Thus, we obtain the average
BER of DSTBC-OFDM by averaging the conditioned instantaneous BER over
the probability distribution function (PDF) of the equivalent SINR.

Since the equivalent instantaneous interference power is the expected
power of entries in $\mathbf{\Theta}$ for a given $\mathbf{\Lambda}$
and $\bar{\mathbf{\Lambda}}$. From Eq. (\ref{eq:lmda}), the entries
of the diagonal matrices $\mathbf{\Lambda}$ and $\bar{\mathbf{\Lambda}}$
correspond to the DFT of the multipath channel impulse response whose
$L$ paths follow a zero-mean Gaussian distribution \cite{Torabi2007}.
Hence, the diagonal matrices $\mathbf{\Lambda}$ and $\bar{\mathbf{\Lambda}}$
follow a zero-mean Gaussian distribution with unit variance. 

In addition, the equivalent interference matrix $\mathbf{\Theta}$
is the sum of two matrices which are the product of four independent
Gaussian variables. Hence, we have $E\{{{\mathbf{S}}_{k}}^{H}{{\mathbf{\Lambda}}^{H}}\bar{\Lambda}\mathbf{{\mathbf{\bar{S}}}}_{k+1}{{\mathbf{S}}_{k+1}}^{H}{{\mathbf{\Lambda}}^{H}}\bar{\Lambda}\mathbf{{\mathbf{\bar{S}}}}_{k}\}=\mathbf{0}$.
Thus, the conditional average power of each entry of the matrix $\mathbf{\Theta}$
is given by
\begin{equation}
{{E}_{\Theta}}=\frac{1}{4}E\{Tr({{\mathbf{\Theta}}^{H}}\mathbf{\Theta})\left|\mathbf{\Lambda},\mathbf{\bar{\Lambda}}\right.\}=|{{\alpha}_{r}}{{\beta}_{r}}{{|}^{2}}|{\mathbf{\lambda}}{{|}^{2}}|\mathbf{\bar{\lambda}}{{|}^{2}}
\end{equation}

Similarly, the conditional average signal and noise power for a given
channel realization can be expressed as follows 
\begin{align}
{E_{S}} & =E\left\{ Tr\left(\mathbf{U}_{k+1}^{H}\mathbf{U}_{k+1}\right)\left|\mathbf{\Lambda},\mathbf{\bar{\Lambda}}\right.\right\} \nonumber \\
 & =\frac{|\lambda|^{4}|{{\alpha}_{r}}{{|}^{4}}}{4}E\left\{ Tr\left(\mathbf{U}_{k+1}^{H}\mathbf{U}_{k+1}\right)\right\} =\frac{1}{2}|{{\alpha}_{r}}\lambda{{|}^{4}}\\
{{E}_{v}} & =\frac{1}{4}E\left\{ Tr({\mathbf{V}}_{r}^{H}{\mathbf{V}}_{r})\left|\mathbf{\Lambda},\mathbf{\bar{\Lambda}}\right.\right\} =2|{{\alpha}_{r}}{{|}^{4}}|\lambda{{|}^{2}}{{\sigma}^{2}}
\end{align}

Therefore, the conditional equivalent instantaneous SINR $\eta_{d}$
of $\mathbf{U}_{k+1}$ in the differential decoding metric for a given
channel realization is given by 

\begin{equation}
{\eta_{d}}=\frac{{E_{S}}}{{{E}_{\mathbf{\Theta}}}+{{E}_{v}}}=\frac{|\lambda{{|}^{2}}}{2|\bar{\lambda}{{|}^{2}}\rho+4{{\sigma}^{2}}}\label{eq:etaprecise}
\end{equation}

\end{singlespace}

First, we analyze the asymptotic performance by setting ${\sigma}^{2}\rightarrow0$,
resulting in the asymptotic equivalent SINR

\begin{singlespace}
\begin{equation}
{{\eta}_{d}^{\left(a\right)}}=\underset{\sigma^{2}\rightarrow0}{\mbox{lim}}\:\eta_{d}=\frac{|\lambda{{|}^{2}}}{2|\mathbf{\mathbf{\bar{\lambda}}}{{|}^{2}}\rho}\label{etaasym}
\end{equation}

Since $\lambda$ and $\mathbf{\bar{\lambda}}$ are independent complex
Gaussian random variables, the ratio $\mbox{X}$ of their squared-absolute
values, $\mbox{X}=\frac{\left|\lambda\right|^{2}}{\left|\overline{\lambda}\right|^{2}}$,
follows the F-distribution \cite{degroot2010} with a probability
density function $p(X)$ given by $p(X)=F(x,4,4)$ where $F(x,b,c)=I_{\frac{bx}{bx+c}}\left(\frac{b}{2},\frac{c}{2}\right)$
and $I$ is the regularized incomplete beta function.

It can be proved that $E\left(X\right)=\underset{x}{\int}X\,p(X)\,dX=2$.
Hence, the asymptotic average equivalent SINR is given by
\end{singlespace}

{\small{}
\begin{align}
E\left(\eta_{d}^{\left(a\right)}\right) & =\frac{\left|\alpha_{r}\right|^{2}}{2\left|\beta_{r}\right|^{2}}E\left(\frac{\left|\lambda\right|^{2}}{\left|\overline{\lambda}\right|^{2}}\right)=\frac{\left|\alpha_{r}\right|^{2}}{2\left|\beta_{r}\right|^{2}}E\left(X\right)=\frac{\left|\alpha_{r}\right|^{2}}{\left|\beta_{r}\right|^{2}}\nonumber \\
 & =1/\rho=\mbox{IRR}\label{eeta}
\end{align}
}{\small \par}

\begin{singlespace}
Based on the general relationship between the BER and the instantaneous
SINR ${\eta}$ of an MPSK signal in \cite{Torabi2007}, the average
asymptotic BER (error floor), denoted by $P_{e,a}$, in the presence
of RX-IQI is given by

{\small{}
\begin{equation}
P_{e,a}=\int\limits _{0}^{\infty}{\frac{1}{{{\log}_{2}}M}erfc\left(\sqrt{{{\eta}_{d}^{\left(a\right)}}}\sin\left(\pi/M\right)\right)p({{\eta}_{d}^{\left(a\right)}})d{{\eta}_{d}^{\left(a\right)}}}\label{BER-psk}
\end{equation}
}where $p\left({{\eta}_{d}^{\left(a\right)}}\right)=2\rho F(2\rho x,4,4)$
is the probability distribution function of SINR ${\eta}_{d}^{\left(a\right)}$
given in Eq. (\ref{etaasym}).

We note that $\beta_{r}\ll\alpha_{r}$ since the interference power
from the image subcarrier is much smaller than that of the desired
signal. Hence, the interference can be treated as Gaussian noise \cite{Jafar}
without loss of generality. Thus, the instantaneous interference power
$2|\bar{\lambda}{{|}^{2}}\rho$ in Eq. (\ref{eq:etaprecise}) could
be replaced by its average ($i.e.$ $E\{2|\bar{\lambda}{{|}^{2}}\rho\}$)
and incorporated into the noise term. Since $|\bar{\lambda}|^{2}$
is the sum of two identical and independent distributed (i.i.d) zero-mean
complex Gaussian random variables with unit variance, hence, $|\bar{\lambda}|^{2}$
is given by $|\bar{\lambda}|^{2}={|{\lambda_{1}}}(N-n+2)|^{2}+{|{\lambda}_{2}}(N-n+2)|^{2}$.
Then, the average interference power is given by $E\{2|\bar{\lambda}{{|}^{2}}\rho\}=4\rho$.
Thus, the instantaneous SINR $\eta_{d}$ in Eq. (\ref{eq:etaprecise}),
for the case of $\beta_{r}\ll\alpha_{r}$, becomes a Chi-square random
variable with $4$ degrees of freedom which can be expressed as follows\vspace{-0.2cm}
\end{singlespace}

{\small{}
\begin{equation}
\eta_{d}\left|_{\beta_{r}\ll\alpha_{r}}\right.=\frac{1}{4}\left|\lambda\right|^{2}\left(\rho+\sigma^{2}\right)^{-1}\label{eq:etanonasym}
\end{equation}
}{\small \par}

\begin{singlespace}

\end{singlespace}

From Eq. (\ref{eq:etanonasym}), the BER floor appears roughly at
the SNR level where the RX-IQI interference power, controlled by $\rho,$
overwhelms the noise power $\sigma^{2}$ (we assume 10 times larger),
which means the BER floor approximately appears when the corresponding
SINR ${{\eta}_{\mathrm{floor}}}$ satisfies the following conditions\vspace{-0.1cm}

\begin{singlespace}
{\small{}
\begin{equation}
\begin{aligned} & {{\eta}_{\mathrm{floor}}}\gg1/\rho\\
\rightarrow & {{\eta}_{\mathrm{floor}}}(\mathrm{dB})\approx\mathrm{IRR(dB)+10dB}
\end{aligned}
\label{eq:flooraper}
\end{equation}
}{\small \par}

Let ${\eta}_{\mathrm{ideal}}$ be the equivalent SNR of an IQI-free
DSTBC-OFDM system that has a BER equal to the BER floor $P_{e,a}$,
which is given by\vspace{-0.1cm}

{\small{}
\begin{equation}
{\eta}_{\mathrm{ideal}}(\mathrm{dB})=-10\log_{10}(\rho)=\mathrm{IRR}(\mathrm{dB})\label{eq:flooreqideal}
\end{equation}
}{\small \par}
\end{singlespace}

This indicates that the best BER under RX-IQI equals the BER of an
IQI-free system when SNR is equal to IRR.

\begin{singlespace}
Since the SINR in Eq. (\ref{eq:etanonasym}) is Chi-squared distributed
with 4 degrees of freedom. We calculate BER for any SNR by approximately
evaluating the integral in Eq. (\ref{BER-psk}) in a closed-form as
follows\vspace{-0.4cm}

{\small{}
\begin{align}
P_{e} & \approx0.2\left(1+1.75\frac{(\rho+{\sigma}^{2})^{-1}}{M^{1.9}+1}\right)\nonumber \\
 & =0.2\left(1+1.75\frac{\mbox{SN\ensuremath{R_{eq}}}}{M^{1.9}+1}\right)^{-2}\label{eq:BERclosedform}
\end{align}
}{\small \par}
\end{singlespace}

Note that the interference due to RX-IQI plays the same role as noise
power as shown in Eq.\eqref{eq:BERclosedform}. The term $(\rho+{\sigma}^{2})^{-1}$
could be viewed as an equivalent SNR, denoted by $\mbox{SN\ensuremath{R_{eq}}}$
($\mbox{SN\ensuremath{R_{eq}^{-1}} = \ensuremath{\mbox{SNR}^{-1}} + \ensuremath{\mbox{IRR}^{-1}}}$),
which is the harmonic mean of the SNR ($\mbox{\ensuremath{{\sigma}^{2}}}=\mbox{SNR}^{-1}$)
and the IRR $\left(\mbox{\ensuremath{\rho}}=\mbox{IR\ensuremath{\mbox{R}^{-1}}}\right)$
which is always less than the minimum of the two and is maximized
when both are equal, i.e., $\mbox{SNR}=\mbox{IRR}$.

For the high SNR scenario, in the case of high RX-IQI levels and hence
a low IRR level, the equivalent SNR $SNR_{eq}$ and hence the BER
$P_{e}$ in Eq. \eqref{eq:BERclosedform} will be dominated by the
IRR level, $i.e.$ $\mbox{SN\ensuremath{R_{eq}}}\approx\mbox{IRR \ensuremath{\approx}\ 1/\ensuremath{\rho}}$.
On the other hand, for the IQI-free scenario, the BER $P_{e}$ in
Eq. \eqref{eq:BERclosedform} will be dominated by the SNR level,
$i.e.$ $\mbox{SN\ensuremath{R_{eq}}}\approx\mbox{SNR \ensuremath{\approx}\ 1/\ensuremath{\sigma}}^{2}$
and as SNR increases, Eq. \eqref{eq:BERclosedform} clearly shows
that the diversity order is $2$ as expected. Moreover, for a higher
order constellation $i.e.$ larger $M$, the BER is more sensitive
to both noise and RX-IQI effects.\vspace{0.2cm}

\begin{singlespace}

\subsection{Comparison with Coherent Detection}
\end{singlespace}

\begin{singlespace}
In this subsection, we compare the effect of RX-IQI in differential
detection with its effect in coherent detection where the information
block $\mathbf{\mathbf{U}}_{k}\text{(}n\text{)}$ is directly transmitted
without differential encoding (we remove the index $k$ for notational
simplicity). The received signal block becomes\vspace{-0.2cm}

{\small{}
\begin{equation}
\mathbf{\mathbf{Z'}}_{coh}(n)=\mathbf{A}_{r}\mathbf{\Lambda}(n)\mathbf{\mathbf{U}}\text{(}n\text{)}\text{+}{{\mathbf{B}}_{r}}\mathbf{\bar{\Lambda}}(n)\mathbf{\bar{U}}\text{(}n\text{)}+|{{\alpha}_{r}}|{\mathbf{V}}+|{{\beta}_{r}}|{\mathbf{\bar{V}}}
\end{equation}
}{\small \par}

Assuming that the receiver has perfect channel state information (CSI),
the coherent detection process at the receiver can be expressed as\cite{Alamouti1998}

\begin{equation}
\mathbf{\hat{U}}(n)=\underset{\mathbf{U}}{\mathop{\arg\max}}\,\left\{ {{\mathbf{U}}^{H}}\mathbf{\Lambda}{{(n)}^{H}}\mathbf{\mathbf{Z'}}_{coh}(n)\right\} \label{eq:Ucoh}
\end{equation}
where $\mathbf{\Lambda}{{(n)}^{H}}{\mathbf{\mathbf{Z}'}_{coh}(n)}$
can be approximated as follows

{\small{}
\begin{equation}
\begin{aligned}\mathbf{\Lambda}{{(n)}^{H}}{\mathbf{\mathbf{Z}'}_{coh}(n)}\approx & |{{\alpha}_{r}}||\mathbf{\lambda}(n){{|}^{2}}\mathbf{U}\text{(}n\text{)}\\
 & +|{{\beta}_{r}}|\mathbf{\Lambda}{{(n)}^{H}}\mathbf{\bar{\Lambda}}(n)\mathbf{\bar{U}}\text{(}n\text{)}\\
 & +\mathbf{\Lambda}{{(n)}^{H}}{{\mathbf{A}}_{r}}\mathbf{V}(n)
\end{aligned}
\end{equation}
}{\small \par}

Following the same analysis in as Section \ref{sub:ber}, the conditional
equivalent instantaneous SINR $\eta_{c}$ for the coherent detection
of $\mathbf{U}\text{(}n\text{)}$ for a given channel realization
is given by \vspace{-0.1cm}

{\small{}
\begin{equation}
{{\eta}_{c}}=\frac{|\lambda{{|}^{2}}}{\bar{|\lambda}{{|}^{2}}\rho+2{{\sigma}^{2}}}\label{eq:coh}
\end{equation}
}{\small \par}

Comparing ${\eta}_{c}$ with $\eta_{d}$ in Eq. (\ref{eq:etaprecise}),
in case of IQI-free system ($i.e.$ ${\alpha}_{r}=1$ and ${\beta}_{r}=0$)
the noise power is half its value in differential decoding, which
leads to a 3dB loss in SNR as observed in \cite{Hughes2000}. In the
presence of RX-IQI, a doubled interference power will be introduced
to the detection because both the previous block and current block
are affected by interference due to RX-IQI in differential detection,
while in coherent detection we assume perfect CSI in the IQI-free
system. Thus, based on our previous analysis of RX-IQI in differential
systems, the BER of coherent detection can be obtained by setting
both the noise power ${\sigma}^{2}$ and IQI interference power $\rho$
to half of their values in a differential detection. Equivalently,
the performance gap between differential and coherent STBC detection
consists of a 3dB loss in SNR and also a 3dB loss in IRR ($\mbox{IRR = 1/\ensuremath{\rho}}$)
in the differential system. Since BER is sensitive to RX-IQI, the
equivalent SNR degradation caused by a 3dB loss in IRR of RX-IQI is
appreciable. Hence, in the presence of high RX-IQI, the performance
gap between coherent and differential system could be much larger
than 3dB.\vspace{0.24cm}
\end{singlespace}

\begin{singlespace}

\section{\label{sec:compensateion}Estimation and Compensation Algorithm for
RX-IQI in DSTBC-OFDM}
\end{singlespace}

\begin{singlespace}
The frequency-domain RX-IQI-distorted received signals in Eqs. (\ref{eq:freDiff2})
and (\ref{eq:freqDiff-1}) can be expressed in the widely-linear equivalent
form shown in Eqs. (\ref{WLTRIQ-2}) and (\ref{WLTRIQ-1-1}) below
where $\mathbf{S}_{k+1}\text{(}n\text{)}=\mathbf{S}_{k}\text{(}n\text{)\ensuremath{\mathbf{U}_{k+1}\text{(}}n\text{)}}$.
\vspace{-0.3cm}
\end{singlespace}

{\small{}
\begin{align}
\left[\begin{matrix}\mathbf{\mathbf{Z'}}_{k}\left(n\right)\\
\mathbf{\mathbf{\bar{Z}'}}_{k}\left(n\right)
\end{matrix}\right] & =\underbrace{\left[\begin{matrix}{{\mathbf{A}}_{r}}\mathbf{\Lambda}(n) & {{\mathbf{B}}_{r}}\mathbf{\overline{\Lambda}}(n)\\
\mathbf{B}_{r}^{*}\mathbf{\Lambda}(n) & \mathbf{A}_{r}^{*}\mathbf{\overline{\Lambda}}(n)
\end{matrix}\right]}_{\Phi\left(n\right)}\left[\begin{matrix}\mathbf{S}_{k}\text{(}n\text{)}\\
\mathbf{\overline{S}}_{k}\text{(}n\text{)}
\end{matrix}\right]\nonumber \\
 & +\left[\begin{matrix}{{\mathbf{A}}_{r}}\mathbf{V}_{k}(n)\text{+}{{\mathbf{B}}_{r}}\mathbf{\overline{V}}_{k}(n)\\
\mathbf{A}_{r}^{*}\mathbf{\overline{V}}_{k}(n)\text{+}\mathbf{B}_{r}^{*}\mathbf{V}_{k}(n)
\end{matrix}\right]\label{WLTRIQ-2}
\end{align}
}{\small \par}

{\small{}
\begin{align}
\left[\begin{matrix}\mathbf{\mathbf{Z'}}_{k+1}\left(n\right)\\
\mathbf{\mathbf{\bar{Z}'}}_{k+1}\left(n\right)
\end{matrix}\right] & =\underbrace{\left[\begin{matrix}{{\mathbf{A}}_{r}}\mathbf{\Lambda}(n) & {{\mathbf{B}}_{r}}\mathbf{\overline{\Lambda}}(n)\\
\mathbf{B}_{r}^{*}\mathbf{\Lambda}(n) & \mathbf{A}_{r}^{*}\mathbf{\overline{\Lambda}}(n)
\end{matrix}\right]}_{\Phi\left(n\right)}\left[\begin{matrix}\mathbf{S}_{k+1}\text{(}n\text{)}\\
\mathbf{\overline{S}}_{k+1}\text{(}n\text{)}
\end{matrix}\right]\nonumber \\
 & +\left[\begin{matrix}{{\mathbf{A}}_{r}}\mathbf{V}_{k+1}(n)\text{+}{{\mathbf{B}}_{r}}\mathbf{\overline{V}}_{k+1}(n)\\
\mathbf{A}_{r}^{*}\mathbf{\overline{V}}_{k+1}(n)\text{+}\mathbf{B}_{r}^{*}\mathbf{V}_{k+1}(n)
\end{matrix}\right]\label{WLTRIQ-1-1}
\end{align}
}{\small \par}

\begin{singlespace}
Hence, the detected $2\times2$ STBC transmitted data matrices $\mathbf{\hat{S}}_{k}(n)$
and $\mathbf{\hat{\bar{S}}}_{k}\text{(}n\text{)}$ corresponding to
the $\left(2k+1\right)$-th, $\left(2k+2\right)$-th OFDM symbols
using the widely-linear model in Eq. (\ref{WLTRIQ-2}) can be expressed
as follows\vspace{-0.2cm} 
\begin{equation}
\left[\begin{matrix}\mathbf{\hat{S}}_{k}(n)\\
\mathbf{\hat{\bar{S}}}_{k}\text{(}n\text{)}
\end{matrix}\right]=\underbrace{\left[\begin{matrix}{{\mathbf{\Gamma}}_{11}}\text{(}n\text{)} & {{\mathbf{\Gamma}}_{12}}\text{(}n\text{)}\\
{{\mathbf{\Gamma}}_{21}}\text{(}n\text{)} & {{\mathbf{\Gamma}}_{22}}\text{(}n\text{)}
\end{matrix}\right]}_{\mathbf{\Gamma}\text{(}n\text{)}}\left[\begin{matrix}\mathbf{\mathbf{Z'}}_{k}\left(n\right)\\
\mathbf{\mathbf{\bar{Z}'}}_{k}\left(n\right)
\end{matrix}\right]
\end{equation}

\end{singlespace}

In the absence of noise, the transmitted symbol is perfectly recovered
when $\mathbf{\Gamma}(n)={{\mathbf{\Phi}}^{-1}}(n)$. However, since
CSI is unknown in DSTBC-OFDM, it is not possible to invert $\Phi$.
Thus, a new strategy should be used to estimate and compensate IQI
in DSTBC-OFDM. Unlike coherent systems, we do not need to recover
the transmitted signal, instead, we only need to ensure that the differential
relationship in Eq. (\ref{difsu}) is still satisfied by the adjacent
data blocks. However, by examining Eqs. (\ref{eq:freDiff2}) and (\ref{eq:freqDiff-1}),
we find that the differential relationship no longer holds in the
presence of RX-IQI even without noise, $i.e.$\vspace{0.15cm} 
\begin{equation}
\left[\begin{matrix}\mathbf{\mathbf{Z'}}_{k+1}\left(n\right)\\
\mathbf{\mathbf{\bar{Z}'}}_{k+1}\left(n\right)
\end{matrix}\right]\ne\left[\begin{matrix}\mathbf{\mathbf{Z'}}_{k}\left(n\right)\mathbf{U}_{k+1}\text{(}n\text{)}\\
\mathbf{\mathbf{\bar{Z}'}}_{k}\left(n\right)\mathbf{\overline{U}}_{k+1}\text{(}n\text{)}
\end{matrix}\right]
\end{equation}

It can be verified that the necessary condition to enforce this relationship
is to maintain the following relationship 
\begin{equation}
\mathbf{\Gamma}\text{(}n\text{)}{\mathbf{\Phi}\text{(}n\text{)}}=\left[\begin{matrix}{{\mathbf{H}}_{1}} & {{\mathbf{0}}_{2\times2}}\\
{{\mathbf{0}}_{2\times2}} & {{\mathbf{H}}_{2}}
\end{matrix}\right]\label{Dig}
\end{equation}
where ${{\mathbf{H}}_{i}}$ ($i=1,2$) are non-unique Alamouti matrices
which are related to the channel and RX-IQI parameters. Based on the
necessary conditions in Eq. (\ref{Dig}), the following relations
should hold\vspace{-0.12cm} 
\begin{equation}
\left\{ \begin{aligned}{\mathbf{\Gamma}}_{11}{{\mathbf{B}}_{r}}\mathbf{\bar{\Lambda}}(n)+{{\mathbf{\Gamma}}_{12}}{\mathbf{A}_{r}^{*}}^{\text{}}\mathbf{\bar{\Lambda}}(n)=0\\
{{\mathbf{\Gamma}}_{21}}{{\mathbf{A}}_{r}}{\mathbf{\Lambda}(n)}+{{\mathbf{\Gamma}}_{22}}{\mathbf{B}_{r}^{*}}^{\text{}}{\mathbf{\Lambda}(n)}=0
\end{aligned}
\right.\label{qrs}
\end{equation}

Since any non-zero matrix $\mathbf{\Gamma}\text{(}n\text{)}$ which
satisfies the relations in Eq. (\ref{qrs}) satisfies the differential
property, we set ${{\mathbf{\Gamma}}_{11}}={{\mathbf{\Gamma}}_{22}}=\mathbf{I}$
for simplicity. Thus, we only need to have the following condition
met to satisfy the differential property 
\begin{equation}
{{\mathbf{\Gamma}}_{12}}={{\mathbf{\Gamma}}_{21}^{*}}=-{\mathbf{B}_{r}}(\mathbf{A}_{r}^{*})^{-1}\triangleq\mathbf{\Gamma}=\left[\begin{array}{cc}
\mathbf{\gamma} & 0\\
0 & \gamma^{*}
\end{array}\right]\label{gammac}
\end{equation}

\begin{singlespace}
Hence, the recovered transmitted data matrices $\mathbf{\hat{S}}_{k}(n)$
and $\mathbf{\hat{\bar{S}}}_{k}\text{(}n\text{)}$ can be expressed
as follows\vspace{-0.3cm}

\begin{equation}
\left[\begin{matrix}\mathbf{\hat{S}}_{k}(n)\\
\mathbf{\hat{\bar{S}}}_{k}\text{(}n\text{)}
\end{matrix}\right]=\left[\begin{array}{cc}
\mathbf{I} & \mathbf{\mathbf{\Gamma}}\\
\mathbf{\Gamma}^{*} & \mathbf{I}
\end{array}\right]\left[\begin{matrix}\mathbf{\mathbf{Z'}}_{k}\left(n\right)\\
\mathbf{\mathbf{\bar{Z}'}}_{k}\left(n\right)
\end{matrix}\right]
\end{equation}

Similarly, the recovered $2\times2$ STBC transmitted data matrices
$\mathbf{\hat{S}}_{k+1}(n)$ and $\mathbf{\hat{\bar{S}}}_{k+1}\text{(}n\text{)}$
corresponding to the $\left(2\left(k+1\right)+1\right)$, $\left(2\left(k+1\right)+2\right)$
OFDM symbols using Eq. (\ref{WLTRIQ-1-1}) can be expressed as follows\vspace{-0.4cm}

\begin{equation}
\left[\begin{matrix}\mathbf{\hat{S}}_{k+1}(n)\\
\mathbf{\hat{\bar{S}}}_{k+1}\text{(}n\text{)}
\end{matrix}\right]=\left[\begin{array}{cc}
\mathbf{I} & \mathbf{\mathbf{\Gamma}}\\
\mathbf{\Gamma}^{*} & \mathbf{I}
\end{array}\right]\left[\begin{matrix}\mathbf{\mathbf{Z'}}_{k+1}\left(n\right)\\
\mathbf{\mathbf{\bar{Z}'}}_{k+1}\left(n\right)
\end{matrix}\right]
\end{equation}

Since, there is no training phase in differential transmission, the
estimation of the parameter $\mathbf{\mathbf{\Gamma}}$, or equivalently
$\gamma$ in \eqref{gammac}, can only be done based on the received
data. We propose a decision-directed method to estimate the compensation
parameter $\mathbf{\mathbf{\Gamma}}$. Based on a least-squares angle
estimator, $\mathbf{\mathbf{\Gamma}}$ can be estimated as follows\vspace{-0.5cm}
\end{singlespace}

\begin{align*}
\Gamma & =\mbox{arg }\mbox{\ensuremath{\underset{\Gamma}{\mbox{min}}}}E\biggl\{\left|\mathbf{\mathbf{Z'}}_{k+1}(n)+{\mathbf{\mathbf{\Gamma}}}\mathbf{\mathbf{\bar{Z}'}}_{k+1}(n)\right.\\
 & -\left.\left(\mathbf{\mathbf{Z'}}_{k}(n)+\mathbf{\mathbf{\Gamma}}\mathbf{\mathbf{\bar{Z}'}}_{k}(n)\right)\mathbf{U}_{k+1}(n)\right|^{2}\biggr\}
\end{align*}

\begin{align}
 & =\mbox{arg }\mbox{\ensuremath{\underset{\Gamma}{\mbox{min}}}}E\Biggl\{\left|\underbrace{\mathbf{\mathbf{Z'}}_{k+1}(n)-\mathbf{\mathbf{Z'}}_{k}(n)\mathbf{U}_{k+1}(n)}_{\mathbf{\Xi}_{k}(n)}\right.\nonumber \\
 & +\left.\mathbf{\mathbf{\Gamma}}\underbrace{(\mathbf{\mathbf{\bar{Z}'}}_{k+1}(n)-\mathbf{\mathbf{\bar{Z}'}}_{k}(n)\mathbf{U}_{k+1}(n))}_{\mathbf{\Delta}_{k}(n)}\right|^{2}\Biggr\}\label{49-1}
\end{align}

\begin{singlespace}
Since the matrices $\mathbf{\Xi}_{k}(n)$ and $\mathbf{\Delta}_{k}(n)$
defined above enjoy the orthogonal Alamouti structure, the estimation
can be simplified by considering only the $1^{st}$ column of $\mathbf{\Xi}_{k}(n)$
and $\mathbf{\Delta}_{k}(n)$. Thus, Eq. (\ref{49-1}) can be simplified
as follows\vspace{-0.3cm}

\begin{align}
{\mathbf{\mathbf{\Gamma}}} & =\arg\ \underset{{\mathbf{\mathbf{\Gamma}}}}{\mathop{\min}}\,E\Biggl\{{\left|\left[\begin{matrix}[\mathbf{\Xi}_{k}(n)]{_{1,1}} & [\mathbf{\Xi}_{k}{(n)}]{_{2,1}}\end{matrix}\right]^{T}\right.}\nonumber \\
 & \left.+{\mathbf{\mathbf{\Gamma}}}\left[\begin{matrix}{{[\mathbf{\Delta}_{k}(n)]}_{1,1}} & {{[\mathbf{\Delta}_{k}{(n)}]}_{2,1}}\end{matrix}\right]^{T}\right|^{2}\biggr\}\label{eq:gammasimple}
\end{align}
where $[\mathbf{\Xi}_{k}(n)]{_{1,1}}$ and $[\mathbf{\Xi}_{k}{(n)}]{_{2,1}}$
are the elements of the $1^{st}$ column of the matrix $\mathbf{\Xi}_{k}(n)$.
In addition, $[\mathbf{\Delta}_{k}(n)]{_{1,1}}$ and ${{[\mathbf{\Delta}_{k}{(n)}]}_{2,1}}$
are the elements of the $1^{st}$ column of the matrix $\mathbf{\Delta}_{k}(n)$.
On the other hand, $\Gamma$ has the special structure in Eq.\eqref{gammac}.
Therefore, Eq.\eqref{eq:gammasimple} could be further simplified
to\vspace{-0.3cm}

\begin{align}
{\mathbf{\mathbf{\gamma}}} & =\arg\ \underset{{\mathbf{\mathbf{\gamma}}}}{\mathop{\min}}\,E\Biggl\{{\left|\left[\begin{matrix}[\mathbf{\Xi}_{k}(n)]{_{1,1}} & [\mathbf{\Xi}_{k}^{*}{(n)}]{_{2,1}}\end{matrix}\right]\right.}\nonumber \\
 & \left.+{\mathbf{\mathbf{\mathbf{\mathbf{\gamma}}}}}\left[\begin{matrix}{{[\mathbf{\Delta}_{k}(n)]}_{1,1}} & {{[\mathbf{\Delta}_{k}^{*}{(n)}]}_{2,1}}\end{matrix}\right]\right|^{2}\biggr\}\label{eq:gammasimple-1}
\end{align}

\end{singlespace}

We use the adaptive Least Mean Square (LMS) algorithm to iteratively
estimate $\gamma$. We define \vspace{-0.3cm}

\begin{singlespace}
\begin{align}
{{e}(m)} & \triangleq{\mathbf{\xi}}(n)+{{\mathbf{\gamma}}(m-1)}{\mathbf{\delta}}(n)\\
{{\mathbf{\gamma}}(m)} & \triangleq{{\mathbf{\gamma}}(m-1)}+\mu{{e}(m)}{{\mathbf{\delta}^{H}}(n)}
\end{align}

where ${\mathbf{\gamma}}(m-1)$ is the estimated compensation parameter
$\gamma$ after $m-1$ iterations and the set $({\mathbf{\xi}}(n),{\mathbf{\delta}}(n))$
is chosen from the available set $\left\{ ([\mathbf{\Xi}_{k}(n)]{_{1,1}},{{[\mathbf{\Delta}_{k}(n)]}_{1,1}}),([\mathbf{\Xi}_{k}^{*}{(n)}]{_{2,1}},{{[\mathbf{\Delta}_{k}^{*}{(n)}]}_{2,1}})\right\} $
defined in Eq. (\ref{eq:gammasimple}). In addition, $\mu$ is the
LMS adaptation step size. \vspace{0.3cm}
\end{singlespace}

\begin{singlespace}

\section{\label{sec:numerical}Numerical Results }
\end{singlespace}

\begin{singlespace}
The system parameters are similar to \cite{narasimhan2009digital}.
The transmitter sends 8-PSK modulated symbols over a bandwidth of
5MHz and the operating frequency is 2.5GHz. The number of OFDM subcarriers
is set to 64. The channel models used for slow fading and fast fading
are the ITU Pedestrian channel B (ITU-PB) and the ITU Vehicular channel
A (ITU-VA), respectively. The mobile speed is 5km/h for slow fading
and 200km/h for fast fading, corresponding to maximum Doppler shifts
of 11.6Hz and 463.0Hz, respectively. The RX-IQI parameters ${{\kappa}_{r}}(dB)=20\log(g_{r})=2\mathrm{dB}$
and ${{\phi}_{r}}={{8}^{\circ}}$, resulting a receiver IRR of 16.8dB. 
\end{singlespace}

\begin{figure}
\includegraphics[width=1\columnwidth,height=86in,keepaspectratio]{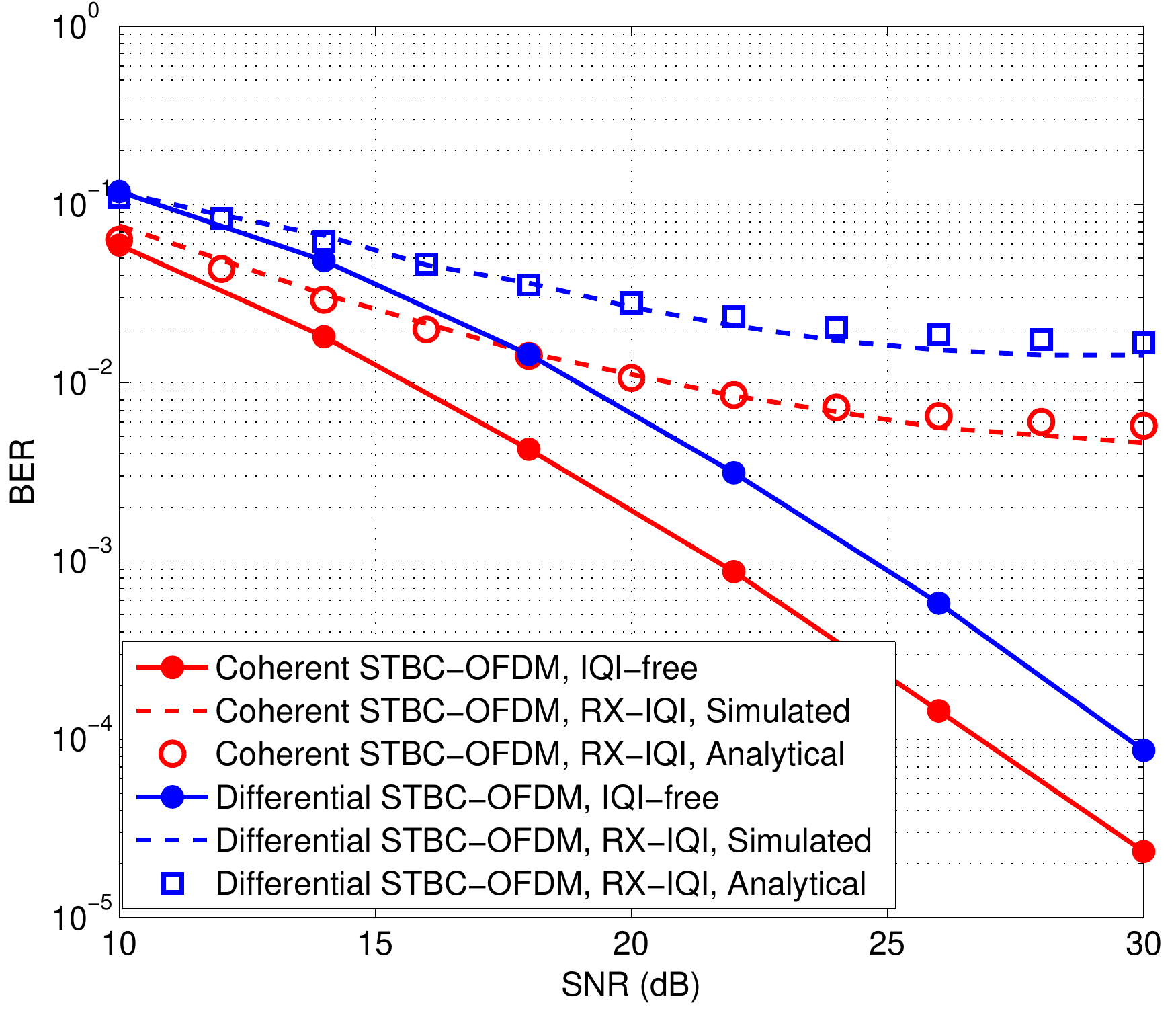}\caption{Comparison of Analytical and Measured BER performance of the DSTBC-OFDM
system under RX-IQI. }

\label{analyticalber}
\end{figure}

Fig. \ref{analyticalber} demonstrates that the analytical BER in
Eq. (\ref{BER-psk}) and simulated BER results match very well. The
approximated closed-form BER in Eq. (\ref{eq:BERclosedform}) is slightly
different from the simulated BER, but it is easy to compute and provides
useful insights about the IQI impact. Also, as expected from Eq. (\ref{eq:coh})
and its interpretations, in the absence of RX-IQI, Fig.\ref{analyticalber}
shows that the gap between coherent detection and differential detection
is roughly 3dB. On the other hand, the BER performance gap increases
drastically and both coherent and differential detection degrade significantly
in BER performance under RX-IQI. It could also be observed that the
BER curve has a floor at high SNR, which, according to our previous
analysis, is caused by the limited SINR even in the absence of noise.
Moreover, the BER floor roughly starts at SNR around $27$ dB which
matches our analysis in Eq. (\ref{eq:flooraper}). The asymptotic
BER under RX-IQI is obtained from Eq. (\ref{etaasym}) and (\ref{BER-psk}),
which according to the simulation, is equal to the BER at $\mbox{SNR}=17.8dB$
in a IQI-free system. This value is only $1$ dB different from our
predicted value due to the model in Eq. (\ref{eq:etanonasym}) being
an approximate model.

\begin{singlespace}

\begin{figure}
\centering \includegraphics[width=1\columnwidth,height=0.86\textheight,keepaspectratio]{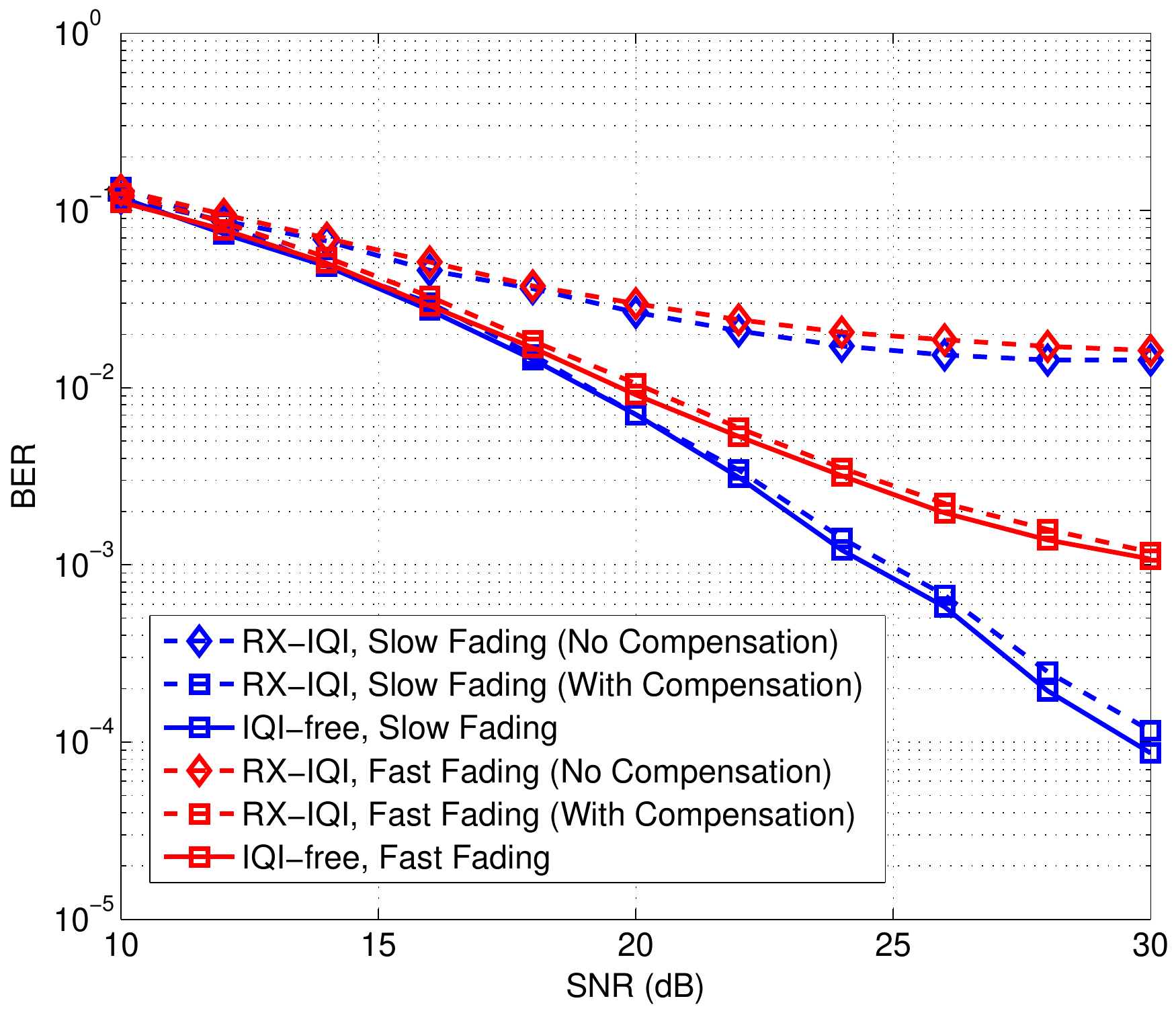}
\caption{BER performance of the DSTBC-OFDM system under RX-IQI in slow and
fast fading channels \label{rx}}
\end{figure}

The performance of RX-IQI compensation is presented in Fig. \ref{rx}.
We evaluate the performance in both fast-fading and slow-fading channels.
Fig. \ref{rx} shows that our proposed compensation algorithm effectively
mitigates RX-IQI. A performance degradation is observed in the fast-fading
channel even without IQI since the fast-varying channel does not satisfy
the quasi-static property assumed by differential STBC. However, since
the RX-IQI compensation matrix does not change with the channel, the
compensation is effective in both slow and fast fading channel scenarios
and the degradation caused by RX-IQI is almost eliminated.

\vspace{0.2cm}
\end{singlespace}

\begin{singlespace}

\section{\label{sec:conc}Conclusion}
\end{singlespace}

\begin{singlespace}
In this paper, we analyzed the impact of RX-IQI on the BER of DSTBC-OFDM
systems. We quantified analytically the BER floor due to RX-IQI and
demonstrated its accuracy by simulations. In addition, a compensation
scheme for DSTBC-OFDM system is proposed and demonstrated to effectively
mitigate the performance degradation caused by RX-IQI in both slow
and fast fading channels.
\end{singlespace}

\section*{Acknowledgment}

\vspace{-0.1cm}This work was done while Lei Chen was a visiting PhD
student at University of Texas at Dallas and his work is supported
in part by the scholarship from China Scholarship Council (CSC). The
work of A. Helmy and N. Al-Dhahir was made possible by NPRP grant
\#NPRP 8-627-2-260 from the Qatar National Research Fund (a member
of Qatar Foundation). The statements made herein are solely the responsibility
of the authors.

\vspace{-0.2cm}\bibliographystyle{IEEEtran}


\end{document}